# Disproof of a widely-accepted mathematical conjecture

**Changbiao Wang**[*]

ShangGang Group, 70 Huntington Road, Apartment 11, New Haven, CT 06512, USA

A mathematical conjecture is successfully identified, which is used for relativistic analysis of dielectric Einstein-box thought experiment in a Letter (Ramos, Rubilar, and Obukhov, Phys. Lett. A 375, 1703 (2011)), where the authors conjecture (without any citations) that, the symmetry and divergence-less property of a Lorentz 4-tensor is a sufficient condition for the time-column space integrals to constitute a Lorentz 4-vector. This mathematical conjecture has been thought to be "a mathematical fact the validity of which was shown well" in textbooks. However in this paper, we indicate that this conjecture has never been proved mathematically. By enumerating a counterexample, we find that this mathematical conjecture is flawed, and it is not persuasive to use a flawed mathematical conjecture as a starting point to resolve Abraham-Minkowski controversy over light momentum in a dielectric medium. We also indicate that this flawed mathematical conjecture is actually a widely-accepted conjecture in the dynamics of relativity in textbooks for many decades. To eliminate a misunderstanding of this flawed conjecture in the community, we provide a detailed elucidation of why Møller's mathematical statement, also called "Møller's version of von Laue's theorem", only defines a trivial *zero* 4-vector for an electromagnetic stress-energy Lorentz 4-tensor.

## I. INTRODUCTION

The momentum of light in a dielectric medium is a fundamental problem in physics, so-called Abraham-Minkowski controversy. To resolve this controversy, various postulates and assumptions, or even mathematical conjectures are invoked. For example, Mansuripur and Zakharian postulate that Poynting vector represents the electromagnetic (EM) power flow in any system of materials, and claim that the Abraham momentum is "the sole electromagnetic momentum in any system of materials distributed throughout the free space" [1]; Brevik assumes that Abraham force is the correct EM force in a medium, and claims that Abraham force "simply fluctuates out when averaged over an optical period in a stationary beam", but "it is in principle measurable" [2]. However both Mansuripur-Zakharian postulate and Brevik's assumption are challenged based on the momentum-energy conservation law when a plane wave propagates in a uniform medium [3,4].

In a beautiful Letter aimed to resolve Abraham-Minkowski controversy, a relativistic analysis is given of dielectric Einstein-box thought experiment (also called "Balazs thought experiment") [5]. The authors conjecture (without any citations) that the symmetry and divergence-less property of a Lorentz 4-tensor is a sufficient condition for the time-column space integrals to constitute a Lorentz 4-vector. This conjecture is thought to be "a mathematical fact the validity of which was shown well" in textbooks. In this paper, we would like to indicate that this conjecture has never been proved mathematically. By enumerating a counterexample, this mathematical conjecture is disproved. The disproof is given below.

After a total tensor is constructed, Ramos, Rubilar, and Obukhov declare a 4-momentum vector based on an implicit assumption, as shown in Fig. 1. From Fig. 1, we can see that the authors implicitly assume that if the total tensor $T_\mu^{\,\nu}$ is *symmetric* and *divergence-less* ($\partial_\nu T_\mu^{\,\nu} = 0$ due to $J_{\text{ext}}^\nu = 0$), the time-column space integrals $P_\mu = \int_{V'} T_\mu^{\,0} dV$ must constitute a Lorentz 4-vector.

If putting aside the physical explanations assigned by the authors, the above Ramos-Rubilar-Obukhov implicit assumption is equivalent to a mathematical conjecture, as stated below.

> *Mathematical conjecture:* If $\Theta^{\mu\nu}(\mathbf{x},t)$ defined in $V$ is a Lorentz 4-tensor, of which all the elements have first-order partial derivatives with respect to time-space coordinates $X^\mu = (\mathbf{x}, ct)$, and it is symmetric ($\Theta^{\mu\nu} = \Theta^{\nu\mu}$) and divergence-less ($\partial_\mu \Theta^{\mu\nu} = 0 \Leftrightarrow \partial_\nu \Theta^{\mu\nu} = 0$), then the time-row (column) space integrals (assumed to be convergent)
>
> $$P^\nu = \int_V \Theta^{4\nu} d^3x \; \left( = \int_V \Theta^{\nu 4} d^3x \right) \tag{1}$$
>
> constitute a Lorentz 4-vector.

---

[*] changbiao_wang@yahoo.com

> T. Ramos et al. / Physics Letters A 375 (2011) 1703–1709
>
> The total tensor is symmetric and satisfies the following energy–momentum balance equation,
>
> $$\partial_\nu T_\mu{}^\nu - F_{\mu\nu} J^\nu_{ext} = 0, \quad (4)$$
>
> where the 4-vector $J^\nu_{ext}$ describes the external charge and current densities which do not belong to the dielectric fluid. If $J^\nu_{ext} = 0$ energy–momentum tensor of the complete system is conserved and we have a closed system.
>
> If we choose a volume $V'$ big enough so that it encloses the pulse and the slab until the pulse leaves the slab from the other side, then we can integrate the conservation equation and obtain that the total 4-momentum $\mathcal{P}_\mu := (E, -\mathbf{p})$ of the whole system, defined as
>
> $$\mathcal{P}_\mu := \int_{V'} T_\mu{}^0 \, dV, \quad (5)$$
>
> is a conserved, i.e. time independent, quantity. We will use this

Fig. 1. Copied paragraphs to show the basic implicit assumption made by Ramos, Rubilar, and Obukhov in [5]. This assumption is equivalent to a mathematical conjecture: If a Lorentz 4-tensor is symmetric and divergence-less, then the space integrals of time-column elements constitute a Lorentz 4-vector. Note that it is not easy to identify this conjecture, because the authors did not provide any citations, and the authors did not clearly explain why they can obtain "the total 4-momentum" just by integrating "the conservation equation".

As we know, the correctness of a mathematical conjecture cannot be legitimately affirmed by enumerating *specific* examples, no matter how many; however, it can be directly negated by finding specific examples, even only one. In the following, given is such a counterexample that disproves the above mathematical conjecture.

## II. COUNTEREXAMPLE

We will show that a charged metal sphere in free space is the right counterexample to disprove the mathematical conjecture by Ramos, Rubilar, and Obukhov.

From the potential 4-vector $A^\mu$ [6], we have the field-strength 4-tensor, given by $F^{\mu\nu} = \partial^\mu A^\nu - \partial^\nu A^\mu$, or

$$F^{\mu\nu} = \begin{pmatrix} 0 & -B_z & B_y & E_x/c \\ B_z & 0 & -B_x & E_y/c \\ -B_y & B_x & 0 & E_z/c \\ -E_x/c & -E_y/c & -E_z/c & 0 \end{pmatrix}. \quad (2)$$

The Maxwell equations $[\nabla \times \mathbf{H} - \partial(c\mathbf{D})/\partial(ct), \nabla \cdot (c\mathbf{D})] = (\mathbf{J}, c\rho)$ can be written as $\partial_\mu G^{\mu\nu} = J^\nu$ [6], where

$$G_{\mu\nu} = g_{\mu\sigma} G^{\sigma\lambda} g_{\lambda\nu} = \begin{pmatrix} 0 & -H_z & H_y & -D_x c \\ H_z & 0 & -H_x & -D_y c \\ -H_y & H_x & 0 & -D_z c \\ D_x c & D_y c & D_z c & 0 \end{pmatrix}, \quad (3)$$

with $g^{\mu\nu} = g_{\mu\nu} = diag(-1,-1,-1,+1)$ being the Minkowski metric. The Lorentz covariant Minkowski electromagnetic (EM) stress-energy 4-tensor is defined as

$$\Pi^{\mu\nu} = g^{\mu\sigma}G_{\sigma\lambda}F^{\lambda\nu} + \frac{1}{4}g^{\mu\nu}G_{\sigma\lambda}F^{\sigma\lambda}, \quad \text{or} \quad \Pi^{\mu\nu} = \begin{pmatrix} \vec{\mathbf{T}}_M & c\mathbf{g}_A \\ c\mathbf{g}_M & W_{em} \end{pmatrix}, \quad (4)$$

where $\mathbf{g}_A = \mathbf{E} \times \mathbf{H}/c^2$ is the Abraham momentum, $\mathbf{g}_M = \mathbf{D} \times \mathbf{B}$ is the Minkowski momentum, $W_{em} = 0.5(\mathbf{D} \cdot \mathbf{E} + \mathbf{B} \cdot \mathbf{H})$ is the EM energy density, and $\vec{\mathbf{T}}_M = -\mathbf{DE} - \mathbf{BH} + \vec{\mathbf{I}}\,0.5(\mathbf{D} \cdot \mathbf{E} + \mathbf{B} \cdot \mathbf{H})$ is the Minkowski stress tensor, with $\vec{\mathbf{I}}$ the unit tensor. $G_{\sigma\lambda}F^{\sigma\lambda} = 2(\mathbf{B} \cdot \mathbf{H} - \mathbf{D} \cdot \mathbf{E})$ is a Lorentz invariant.

From Eq. (4) we have

$$\partial_\mu \Pi^{\mu\nu} = F^{\mu\nu}J_\mu + \left(G^\mu_\lambda \partial_\mu F^{\lambda\nu} + \frac{1}{4}\partial^\nu G_{\sigma\lambda}F^{\sigma\lambda}\right). \quad (5)$$

Suppose that there is a charged metal sphere in free space, with a radius of $R \neq 0$ and a charge of $Q \neq 0$, and the metal sphere is made from perfect conductor so that the EM fields within the sphere ($r < R$) are equal to zero. In free space, $\mathbf{D} = \varepsilon_0 \mathbf{E}$ and $\mathbf{B} = \mu_0 \mathbf{H}$ hold, where $\varepsilon_0$ and $\mu_0$ are the permittivity and permeability constants in vacuum, respectively. Thus for the charged metal sphere, observed in the sphere-rest frame, we have $\mathbf{DE} = \mathbf{ED}$ ($\neq 0$), $\mathbf{BH} = \mathbf{HB}$ ($= 0$), and $\mathbf{g}_A = \mathbf{g}_M$ ($= 0$) holding in the field distribution region ($R < r < +\infty$), leading to $\Pi^{\mu\nu} = \Pi^{\nu\mu}$ from Eq. (4). On the other hand, (a) there are no current ($\mathbf{J} = 0$) and no charge ($\rho = 0$) in the region ($R < r < +\infty$), leading to $J_\mu = (-\mathbf{J}, c\rho) = 0$; (b) the region ($R < r < +\infty$) is filled with a medium of *vacuum*, leading to $G^\mu_\lambda \partial_\mu F^{\lambda\nu} + 0.25\partial^\nu G_{\sigma\lambda}F^{\sigma\lambda} = 0$ [7]. Thus we have $J_\mu = 0$ and $G^\mu_\lambda \partial_\mu F^{\lambda\nu} + 0.25\partial^\nu G_{\sigma\lambda}F^{\sigma\lambda} = 0 \Rightarrow \partial_\mu \Pi^{\mu\nu} = 0$ from Eq. (5). To put it simply, for the charged metal sphere we have:

(i) $\Pi^{\mu\nu}$ is symmetric ($\Pi^{\mu\nu} = \Pi^{\nu\mu}$), and

(ii) $\Pi^{\mu\nu}$ is divergence-less ($\partial_\mu \Pi^{\mu\nu} = 0$),

satisfying the sufficient condition required by Ramos-Rubilar-Obukhov mathematical conjecture. According to the conjecture or Eq. (1),

$$P^\nu = \int_V \Pi^{4\nu} d^3x = \left(\int_V c\mathbf{g}_M d^3x, \int_V W_{em} d^3x\right) \quad (6)$$

must be a Lorentz 4-vector, where $\int_V c\mathbf{g}_M d^3x = \int_V c\mathbf{g}_A d^3x = \int_V [(\mathbf{E} \times \mathbf{H})/c]d^3x = 0$ and $\int_V W_{em} d^3x = \int_V 0.5(\mathbf{D} \cdot \mathbf{E} + \mathbf{B} \cdot \mathbf{H})d^3x = Q^2/(8\pi\varepsilon_0 R)$. However it has been shown in general that in free space, the total (Abraham=Minkowski) electromagnetic momentum and energy for the electrostatic field cannot constitute a Lorentz 4-vector; namely $P^\nu = \int_V \Pi^{4\nu} d^3x$ is *not* a Lorentz 4-vector at all [7,8]. Thus the mathematical conjecture by Ramos, Rubilar, and Obukhov is not true.

## III. CONCLUSIONS AND REMARKS

In summary, we have successfully identified a mathematical conjecture implicitly used in the Letter by Ramos, Rubilar, and Obukhov for resolution of Abraham-Minkowski controversy [5]. By enumerating a counterexample we have shown that the symmetry and divergence-less property of a Lorentz 4-tensor $\Theta^{\mu\nu}(\mathbf{x},t)$ is not a sufficient condition for the time-row (column) space integrals $P^\nu = \int_V \Theta^{4\nu} d^3x$ ($= \int_V \Theta^{\nu 4} d^3x$) to constitute a Lorentz 4-vector. Accordingly, we believe that Ramos-Rubilar-Obukhov mathematical conjecture is flawed, and it is not persuasive to use this flawed conjecture as a starting point for relativistic analysis of Einstein-box thought experiment to resolve Abraham-Minkowski controversy [5]. We also believe that Ramos-Rubilar-Obukhov conjecture is actually the Weinberg's conjecture, as indicated in Ref. [7], where both Weinberg's and Landau-Lifshitz conjectures are shown to be flawed.

It should be noted that compared with Weinberg's mathematical conjecture, Ramos-Rubilar-Obukhov conjecture is more difficult to be identified, as seen in Fig. 1. That is because, (a) the authors did not cite any references, and (b) the authors did not clearly tell why they can obtain "the total 4-momentum" just by integrating "the conservation equation".

It is usually argued in the community that Møller provided an *absolutely rigorous* proof of Ramos-Rubilar-Obukhov conjecture. Unfortunately, this is a misunderstanding. Møller's mathematical statement, also called "Møller's version of von Laue's theorem" [7], is completely different from Ramos-Rubilar-Obukhov conjecture. In Møller's statement, the divergence-less plus a "zero-boundary condition" of a tensor is taken as the sufficient condition, while the symmetry is not required. In contrast, in Ramos-Rubilar-Obukhov conjecture, the divergence-less plus a symmetry is taken as the sufficient condition, while no any boundary conditions are imposed.

Møller's *zero-boundary condition* requires that all the tensor elements be equal to zero on the boundary for any time $(-\infty < t < +\infty)$. In the case of an EM stress-energy tensor given by Eq. (4), Møller's zero-boundary condition requires $\Pi^{\mu\nu} = 0$ on the boundary for any time, including $c\mathbf{g}_A = \mathbf{E} \times \mathbf{H}/c = 0$ $\Rightarrow \mathbf{E} \times \mathbf{H} = 0$ on the boundary for any time $(-\infty < t < +\infty)$.

Physically, Møller's sufficient condition is extremely strong and severe, because it requires that (a) within the finite domain *V* of a physical system, there are no any sources ($\partial_\mu \Pi^{\mu\nu} = 0$), and (b) the EM energy never flows through the closed boundary of *V* for any time ($\Pi^{\mu\nu} = 0$ $\Rightarrow \mathbf{E} \times \mathbf{H} = 0$ for $-\infty < t < +\infty$). Thus this physical system is never provided with any EM energy. In terms of energy conservation law, no EM fields can be supported within the domain *V* in such a case, leading to a zero field solution. Thus Møller's mathematical statement only defines a trivial *zero* 4-vector for an EM stress-energy tensor. That is why the application of Møller's mathematical statement is very limited [7].

Nevertheless, one may persist that Ramos-Rubilar-Obukhov mathematical conjecture "is actually not a conjecture, but a mathematical fact the validity of which was shown well" in textbooks. However that is not true. The fact is that, although Ramos-Rubilar-Obukhov or Weinberg's conjecture is a widely-accepted conjecture in the dynamics of relativity in textbooks for many decades [8], its validity has never been confirmed; instead its validity has been proved to be *false* in the present paper and in [7,8].

**References**

[1] M. Mansuripur and A. Zakharian, "Maxwell's macroscopic equations, the energy–momentum postulates, and the Lorentz law of force," Phys. Rev. E **79**, 026608 (2009).

[2] I. Brevik, "Comment on 'Observation of a push force on the end face of a nanometer silica filament exerted by outgoing light'," Phys. Rev. Lett. **103**, 219301 (2009).

[3] C. Wang, "Electromagnetic power flow, Fermat's principle, and special theory of relativity," Optik **126**, 2703–2705 (2015).

[4] C. Wang, "Is the Abraham electromagnetic force physical?" Optik **127**, 2887–2889 (2016).

[5] T. Ramos, G. F. Rubilar, and Y. N. Obukhov, "Relativistic analysis of the dielectric Einstein box: Abraham, Minkowski and total energy-momentum tensors," Phys. Lett. A **375**, 1703-1709 (2011).

[6] J. D. Jackson, *Classical Electrodynamics* (John Wiley & Sons, NJ, 1999), 3rd Edition.

[7] C. Wang, "von Laue's theorem and its applications," Can. J. Phys. **93**, 1470-1476 (2015).

[8] C. Wang, "Self-consistent theory for a plane wave in a moving medium and light-momentum criterion," Can. J. Phys. **93**, 1510-1522 (2015).